\documentclass{aa}
\usepackage{graphics}
\begin{document}

   \thesaurus{ 06 
		(03.13.5; 
		08.01.2;  
		08.09.2 2MASSW~J1145572+231730; 
		08.12.2   
		08.18.1)   
}

   \title{A search for variability in brown dwarfs and L dwarfs}

   \subtitle{}

   \author{C.A.L. Bailer-Jones \& R. Mundt
          }
   \authorrunning{Bailer-Jones \& Mundt}

   \offprints{\\Coryn Bailer-Jones, calj@mpia-hd.mpg.de}

   \institute{Max-Planck-Institut f\"ur Astronomie, K\"onigstuhl 17,
	D-69117 Heidelberg, Germany
             }

   \date{received 16 April 1999; accepted 15 June 1999}

   \maketitle

   \begin{abstract} 

We have undertaken a differential photometric I band search for
variability in three Pleiades brown dwarfs and three very low mass
field L dwarfs.  Analysis of the resultant time series for the
Pleiades brown dwarfs (Teide~1, Calar~3, Roque~11) gives no evidence
for variability (above 99\% confidence) greater than 0.05 mags in any
of these objects on timescales between 25 minutes and 27 hours.
Time series of two of the L dwarfs, 2MASSW~J0913032+184150 and
2MASSW~J1146345+223053, are also consistent with no intrinsic
variability (no significant variability above 0.07 mags for 30 mins $<
\tau <$ 126 hours, and above 0.025 mags for 30 mins $< \tau <$ 75 hours,
for the two objects respectively).  However, the L dwarf
2MASSW J1145572 +231730 shows evidence for variability of
amplitude 0.04 magnitudes with a tentative period of about 7 hours. It
is therefore interesting that Kirkpatrick et
al.~(\cite{kirkpatrick99}) report an H$\alpha$ emission line of
4.2\AA\ equivalent width for this star: If our detection is confirmed
as a rotation period, it would support the link between rotation and
H$\alpha$ emission in very low mass stars.

      \keywords{methods: observational --
		stars: activity --
		stars: low-mass, brown dwarfs --
		stars: individual: 2MASSW J1145572+231730 --
		stars: rotation
               }
   \end{abstract}

%

\section{Introduction}

Brown dwarfs are of great interest to astronomers.  According to one
definition they can be considered to bridge the gap between the lowest
mass stable-hydrogen-burning stars and the Jupiter-like planets which
do not get hot enough to burn deuterium, the most easily ``combusted''
element.  Based on this definition, current models place their mass in
the range $\approx$\,0.015--0.075\,$M_{\odot}$ (D'Antona \&
Mazzitelli~\cite{dantona94}; Burrows et al.~\cite{burrows97}).
However, for single objects mass is not a directly observable
quantity, so brown dwarf status can only be conferred via secondary
indicators.  Models predict that brown dwarfs spend most of their
lives cooling, and all but the youngest and most massive brown dwarfs
have effective temperatures below about 2800\,K.  Of particular
interest, therefore, are the recently discovered compact objects with
atmospheres characteristic of very low temperatures (Kirkpatrick et
al.~\cite{kirkpatrick99}).  Models of these objects suggest 1500\,K~$<
T_{\rm eff} <$~2000\,K, and the newly introduced class {\it L dwarfs}
has been suggested to emphasise the status of these objects as a low
temperature continuation of the M dwarfs. We therefore expect many of
them to be brown dwarfs.

The intrinsic properties of brown dwarfs and L dwarfs are of
astrophysical interest, and the observations and models to date raise
a number of important questions. For example, what is the nature of
their atmospheres? At temperatures below about 2500\,K, models predict
that dust grains start to condense and appear in the atmosphere (e.g.\
Allard et al.~\cite{allard97}).  How does the angular momentum of
these objects evolve? Knowledge of their rotation speeds as a function
of age and mass will help address the problem of the formation and
evolution of these objects.  Do these objects have magnetic activity
and chromospheres?  In higher mass objects, dynamo models give a
relationship between rotation speed and magnetic activity (as
observed, for example, via H$\alpha$ emission). The measurement of
rotation periods and variability amplitudes is important for
addressing these questions.

To date there has been very little work on measuring variability in
brown dwarfs and L dwarfs.  Tinney \& Tolley~(\cite{tinney99}) provide
some evidence for photometric variability in the M9 brown dwarf
\object{LP\,944-20}, but do not derive a period for what they suspect
to be a rotational modulation. Based on spectroscopy, Tinney \&
Reid~(\cite{tinney98}) report $v\sin i$ of about $28 \pm
2$\,kms$^{-1}$ for this object. Mart\'\i n et al.~(\cite{martin98})
give a $v\sin i$ of $13 \pm 1$\,kms$^{-1}$ for the M6 Pleiades brown
dwarf \object{Teide 2}, and Mart\'\i n et al.~(\cite{martin97}) derive
a $v\sin i$ of $20 \pm 10$\,kms$^{-1}$ for the L5 field brown dwarf
\object{DENIS-P J1228.2-1547}.

\begin{table*}
\begin{minipage}{13.7cm}
\caption[]{Properties of brown dwarf and L dwarf targets. 
All data for the 2MASS objects are from Kirkpatrick et al.\ (\cite{kirkpatrick99}).
For these objects $I$ is their $I_{spec}$ value (magnitude evaluated from a flux-calibrated spectrum).
\label{targets}
}
\begin{tabular}{lllllll}
\hline
name        & IAU name               & $I$       & SpT & H$\alpha$ EW & $\ion{Li}{i}\,\lambda$6708 EW \\
            &                        &           &              & \AA\         & \AA\                        \\
\hline
2M0913      & \object{2MASSW J0913032+184150}    & 19.07        & L3     & $< 0.8$           & $< 1.0$      \\
2M1145      & \object{2MASSW J1145572+231730}    & 18.62        & L1.5   & 4.2 $\pm$0.2      & $< 0.4$      \\
2M1146      & \object{2MASSW J1146345+223053}    & 17.62        & L3     & $\leq 0.3$        & 5.1$\pm$0.2  \\
\object{Roque 11}$^a$& \object{RPL J034712+2428.5}$^b$ & 18.75$^b$ & M8$^a$ & 5.8 $\pm$1.0$^a$  &           \\
\object{Teide 1}$^c$ & \object{TPL J034718+2422.5}$^b$ & 18.80$^b$ & M8$^d$ & 3.5--8.6$^e$ & 1.0$\pm$0.2$^f$ \\
\object{Calar 3}$^f$ &               & 18.73$^g$  & M9$^d$ & 6.5$^f$--10.2$^d$\,$\pm$\,1.0 & 1.8$\pm$0.4$^f$ \\
\hline
\end{tabular}

$^a$Zapatero Osorio et al.\ (\cite{zap97a})\\ 
$^b$Zapatero Osorio et al.\ (\cite{zap99})\\ 
$^c$Rebolo et al.\ (\cite{rebolo95})\\
$^d$Mart\'\i n et al.\ (\cite{martin96})\\ 
$^e$several values are given in the literature: 6.1\,$\pm$\,1.0 (Rebolo et al.~\cite{rebolo95});\\ 
\hspace*{1.5ex}4.5\,$\pm$\,1.0 (Rebolo et al.~\cite{rebolo96});
3.5\,$\pm$\,2.0, 3.7\,\,$\pm$\,2.0, 8.6\,$\pm$\,2.0 (Zapatero Osorio et al.~\cite{zap97a})\\
$^f$Rebolo et al.\ (\cite{rebolo96})\\
$^g$Zapatero Osorio et al.\ (\cite{zap97b})\\

\end{minipage}
\end{table*}

In this paper we present the results of a search for variability in
six very low mass (VLM) stars (Table~\ref{targets}).  Three of these
are Pleiades brown dwarfs. Two (\object{Teide~1} and \object{Calar~3})
are confirmed brown dwarfs, based on their radial velocities and
lithium detections.  The third, \object{Roque~11}, is a probable brown
dwarf and cluster member.  The three other objects (2M0913, 2M1145 and
2M1146) are L dwarfs discovered by the 2MASS survey. One of these,
2M1146, is a brown dwarf based on the detection of lithium in its
spectrum. The other two have no or very little lithium, so are
probably either VLM stars (just above the hydrogen burning limit), or
old high mass ($>0.065 M_{\odot}$) brown dwarfs.

The most likely cause of variability would be rotational modulation of
the emitted flux.  In the case of the three L dwarfs, this could be
due to inhomogenous dust or other clouds rotating across the
unresolved stellar disk.  In theory, a modulation could also be
observed on the timescale of formation and dissipation of the clouds,
although this may be longer than the few days duration of our search.
For all objects, modulation could be caused by surface star spots
induced by a magnetic field.

\section{Observations and Processing}

The project was carried out with the 2.2m telescope at Calar Alto,
Spain, using CAFOS and the 1K$\times$1K TEK13c CCD detector. The
observations were obtained over the six nights 08/09 January to 13/14 January
1999 (MJD 2451187.4266 to 2451192.7560). However, due to poor weather
(including frequent fog and poor seeing) and various technical
problems, less than half of the available time could be spent
obtaining data.  The observing procedure was to cycle around
the science fields in turn, taking five minute exposures in the I
band in each field (\object{Teide~1} and \object{Roque~11} could be
observed in the same field).  A number of other fields were included
for calibration purposes. The spacing between observations of a given
field was non-uniform, although was typically 15--30 minutes on a given
night.

The CCD frames were processed using IRAF.  To help remove fringing in
the CCD, a night sky flat was constructed from a clipped combination
of 31 frames of 15 different fields. 

For each science field, reference stars were selected which satisfied
the following conditions:
\begin{itemize}
\item{have a near-Gaussian, near-circular PSF}
\item{are bright, but never saturated}
\item{are isolated from other objects}
\item{are as near as possible to the science object.}
\end{itemize}
The reference stars were not know a priori to lack variability.
About ten reference stars were typically found in each field.  Fluxes
at a range of aperture sizes were then evaluated for each reference
star and science star using the IRAF aperture photometry task {\it
phot}. Sky subtraction assumed a single value for the sky across the
extraction aperture.

During reduction, 2M1146 was suspected to be a binary on account of
the shape of its PSF. This was confirmed by Kirkpatrick et
al. (\cite{kirkpatrick99}) who report a background star of much
earlier spectral type separated by $\approx 1''$ from 2M1146.

\section{Time Series Analysis}

Intrinsic variations in a science star were searched for by looking
for changes in the differential magnitude of the science star, that
is, changes in the magnitude of the science star relative to a number
of reference stars. This makes the observations relatively insensitive
to changes in atmospheric transparency during or between
exposures. For this to be successful, we must be confident that the
reference stars are insignificantly variable themselves. We therefore
adopted the following procedure to test for variability in the
reference stars.  A reference star, $s$, is chosen, and the average
flux of all other reference stars formed.  The relative magnitude of
$s$ is then evaluated with respect to this average. This is done at
every epoch, producing a time series
$m_{s,1},m_{s,2},\ldots,m_{s,t},\ldots$, from which the mean is subtracted
so that $\sum_t m_{s,t} = 0$.  At each epoch the total uncertainty in
the relative magnitude, $\epsilon_{s,t}$, is also estimated (see
below).  We then use a $\chi^2$ test to decide whether the scatter
of these relative magnitudes is consistent with their measurement
uncertainties, using the statistic
\begin{equation}
\chi^2_s = \sum_t \left( \frac{m_{s,t}}{\epsilon_{s,t}} \right)^2 \ \ \ .
\end{equation}
The null hypothesis for this test is that there is no variability in
the time series. If ${\rm P}(\chi^2_s)$ is the probability that the
null hypothesis is true, then the larger $\chi^2_s$ the smaller ${\rm
P}(\chi^2_s)$. A star will fail the test (show evidence for
variability) if ${\rm P}(\chi^2_s)<0.01$, a 99\% confidence interval.
(This would correspond to 2.5$\sigma$ significance if the distribution
were Gaussian.) $\chi^2_s$ is measured for every reference star, and
if any fails the test, the star with the highest $\chi^2$ is rejected
from the set of reference stars.  This whole process (starting from
forming the time series for each star) is iterated until all remaining
reference stars pass the $\chi^2$ test.  These then define the
reference system from which to determine the relative magnitude time
series of the science star.

The point of using more than one reference star to determine the time
series for the science star is that it increases the photometric
precision of the relative magnitudes. Furthermore, any small residual
variability in any one reference star will have a smaller effect on
the average.

The $\chi^2$ test is a good initial test for variability, as it is
sensitive to any type of variability and not just periodic variations.
However, for the test to be successful, the errors, $\epsilon_{s,t}$,
must be accurately determined. Using photon statistics we can evaluate
the ``formal'' errors in any magnitude determination. These are (1)
photon arrival noise from the object, (2) photon arrival noise from
the sky, and (3) imperfect determination of the subtracted sky
level. There are, however, additional ``informal'' errors which need
to be accounted for if we are not to overestimate $\chi^2$. The most
important of these results from imperfect flat fielding.  We have
estimated (from the construction of different flats) the accuracy of
the flat to be $0.5\%$, and have combined this in quadrature with the
formal error sources to give $\epsilon_{s,t}$.  Other informal error
sources include (4) imperfect centering of the aperture and (5) PSF
variations over the frame. These were minimised by using a
sufficiently large aperture. To maximise the signal-to-noise ratio in
aperture photometry, one usually chooses an aperture of size
approximately equal to the HWHM (half-width at half-maximum) of the
PSF (Howell \cite{howell89}). It is not necessary to get all of the
light -- or apply an aperture correction -- with relative photometry,
provided that in a given frame a single-sized aperture is used and all
stars have the same PSF.  When there are PSF variations (as was the
case for our data), a larger aperture must be used. This reduces the
contribution from the informal errors (4\,\&\,5) at the expense of
increasing the formal errors (1--3). Thus the threshold for
variability detections in any science star will be at a higher
magnitude limit -- but more reliable -- than when using a smaller
aperture.  We found that an aperture size of about $4\times$HWHM
reduced the informal errors to well below the $0.5\%$ level. Yet other
sources of error (such as imperfect charge transfer efficiency and
nonlinearity of the CCD flux sensitivity) are also below this level.

After establishing reliable reference stars, the $\chi^2$ test was
applied to the science star.  If it failed this test (gave evidence
for variability), the next stage was to determine whether there was
evidence for this variability being periodic. For this purpose we
evaluated the {\it modified periodogram} described by Scargle
(\cite{scargle82}).  In contrast to the classical periodogram, it can
be shown that the modified periodogram is equivalent to a
least-squares fit (in the time domain) of sinusoids to the data
(Lomb~\cite{lomb76}).  Furthermore, the method allows a simple
determination of the significance of any peak in the periodogram based
on the known noise in the data.\footnote{Note that the {\it
Lomb--Scargle periodogram} described by Press et al.\ (\cite{numrec})
differs from the periodogram of Scargle (\cite{scargle82}) by a
constant factor. Thus the two methods will generally give different
significances for any peak in the periodogram.} We consider a peak to
be significant only at the 99\% level (i.e.\ when the probability, $p$,
that a peak is due to noise is $< 0.01$).  If a significant peak is
found, the time series is phased to the detected period and examined
to see if it shows cyclic variability.

\section{Results}

We now have four criteria which the data must satisfy before a
science star is flagged as showing evidence for periodic variations:
\begin{enumerate}
\item{the science star has a significant $\chi^2_s$ value \\(${\rm P}(\chi^2_s)<0.01$)}
\item{the time series looks plausible}
\item{the science star shows a significant ($p<0.01$) peak in the periodogram}
\item{the phased light curve at the detected period looks reasonable.}
\end{enumerate}

These tests were applied to the six science stars in
Table~\ref{targets}.  All except 2M1145 failed at least one. From the
results of the $\chi^2$ test we can set {\it approximate} limits to
the amount and timescale of intrinsic variability in these five stars
at the epoch of the observations (Table~\ref{limits}).  For each star
in this table (except 2M1145) there is no significant evidence (i.e.\
${\rm P}(\chi^2_s)$ is $>$\,0.01) for variability above $I_{\rm lim}$
on the timescale shown. (This is not the same as saying that we are
99\% confident that there is no variability above $I_{\rm lim}$.
Moreover, larger variability on timescales to which we were not
sensitive could be present.) Strictly, it is only possible to put a
single magnitude limit on a given range of periods if the
uncertainties ($\epsilon_{s,t}$) are the same for all $t$.  This is
not the case for our data due to changes in atmospheric
conditions. Rather than defining different amplitude limits for
different periods, we simply give an approximate limit for all
periods.  The upper limit period shown is the longest time base in our
data. The shortest period is a sort of Nyquist period for the time
series, and is slightly more than twice the shortest separation
between frames. We emphasise that these figures should not be viewed
as strict magnitude/period sensitivity limits.

\begin{table}
\caption[]{Approximate upper limits on intrinsic I band variability.
$I_{\rm lim}$ is defined as the scatter of the RMS (root mean squared)
values in the relative magnitude time series that would have been
required before we obtained significant (${\rm P}(\chi^2_s)<0.01$)
evidence for variability according to the $\chi^2$ test.  For each
star the search was sensitive to variations in approximately
the range shown.  }
\label{limits}
\begin{tabular}{lll}
\hline
Star      & $I_{\rm lim}$ &  period range  \\
\hline
2M0913    & 0.070     &  30 mins to 126.2 hours \\
2M1145	  & 0.035     &  30 mins to 74.8 hours \\
2M1146    & 0.025     &  30 mins to 74.8 hours \\
\object{Roque 11}  & 0.045     &  25 mins to 26.4 hours \\
\object{Teide 1}   & 0.050     &  25 mins to 26.4 hours \\
\object{Calar 3}   & 0.050     &  25 mins to 27.5 hours \\
\hline
\end{tabular}
\end{table}

\begin{figure}
\resizebox{\hsize}{!}{\includegraphics{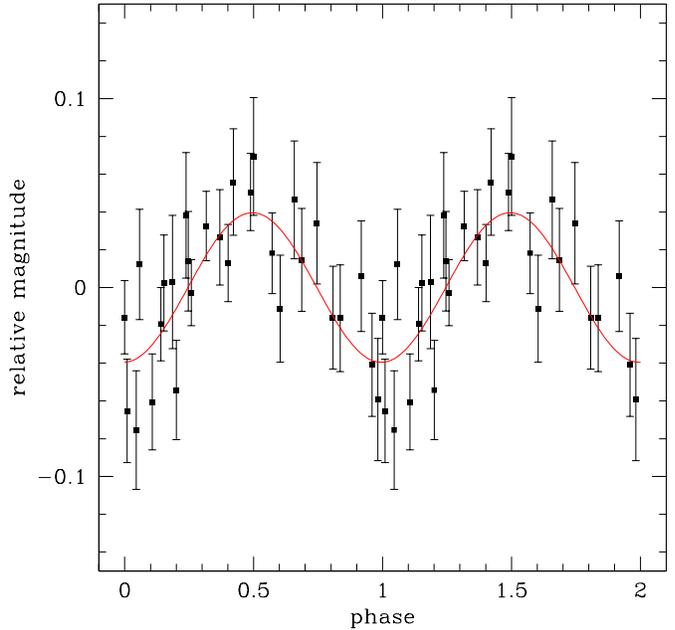}}
\caption{Light curve of the (zero mean) relative magnitude time series 
for the L dwarf 2M1145
phased at the detected period of 7.12 hours.  Two (identical) periods
are shown. The RMS scatter of the variations is 0.038 mags, and the
amplitude of the overplotted least-squares fit sinusoid is 0.040
mags.}
\label{phase}
\end{figure}

The star 2M1145 passed all four tests for variability, and showed a
significant ($p<10^{-4}$) peak in the periodogram at about 7
hours. The phased light curve is shown in Fig~\ref{phase}.  To be
sure that the detected period was not due to a reference star, we
calculated the periodogram of the relative magnitudes of each
reference star (relative to the other reference stars, as described
above). No reference star showed any even marginally significant peak.
Additional checks on the variability detection in 2M1145 were made by
using only a subset of the good reference stars in the reference set,
and using slightly different aperture sizes in the original
photometry. In all cases significant variability was detected
(according to the above four criteria) and the determined periods were
the same to within 1\%.

The RMS (root mean squared) scatter of the relative magnitudes in
Fig~\ref{phase} is 0.038 mags, and the amplitude of a least-squares
fit sinusoid ($A\times\sin[wt + \phi]$) is 0.040 mags. This latter
value may be a slight overestimate of the amplitude of intrinsic
variability on account of noise: the larger the aperture, the more
noise in each measurement, and so the more likely it is that a larger
amplitude is observed.
This does not mean, however, that the detection is just due to noise,
as with a much larger aperture the time series does not meet the
variability criteria described above.  This problem with least-squares
fitting could be overcome using more robust techniques, but we choose
to acquire a more extensive data set before making an improved
determination.  Nonetheless, Fig~\ref{phase} shows evidence
of periodic variation beyond the size of the error bars.

\section{Discussion}

The most plausible explanation for the observed periodic variation in
2M1145 is a rotational modulation of the emitted flux. Assuming a
radius of $0.1 M_{\odot}$ (Burrows et al.~\cite{burrows97}), and rigid
rotation, the period of 7.12 hours implies an equatorial rotation
velocity of 17\,kms$^{-1}$.  This falls in the range of rotation
speeds for 93 field M dwarfs measured by Delfosse et
al.~(\cite{delfosse98}) (all except one with $v\sin i$ in the range
$<$\,2--32\,kms$^{-1}$), but is smaller than the range for 9 Pleiades
M5--M6.5 dwarfs observed by Oppenheimer et al.~(\cite{oppenheimer97})
(37\,$\leq v\sin i \leq$\,65\,kms$^{-1}$).  However, our data does not
give unambiguous evidence for rotation at this speed.  All we can say
for certain is that we have evidence for periodic variability which is
not present in the reference stars, and therefore is probably
intrinsic to 2M1145. But given that we only have 28 points in our time
series spread over several periods, confirmation of this period is
required with a more extensive data set, preferably with smaller error
bars and in more than one filter to provide somewhat independent
measurements of the period.  Observations should also be carried out
over at least two (and preferably three or four) complete periods.  We
also stress that the period determination method assumes that the time
series is stationary, in particular that the period and amplitude of
the variations are constant: any evolution of surface features over
the timescale of the observations would interfere with the
interpretation of the periodogram. Additionally, multiple surface
features may not give rise to a single sinusoidal modulation (and
indeed, other peaks were present in the periodogram).

Given that the observations have only been carried out in a single
filter, we can only speculate about the cause of the modulation in
2M1145.  If its H$\alpha$ emission (Kirkpatrick et
al.~\cite{kirkpatrick99}) can be taken as evidence of magnetic
activity, then the modulation could be the result of magnetically
induced star spots. However, as we have only observed three L dwarfs
we cannot draw any conclusions about the correlation between H$\alpha$
emission and rotation speed at the bottom of the main sequence,
particularly as the amplitude limit on one of the targets (2M0913) is
rather high.  The observed modulations in 2M1145 could alternatively
be the result of inhomogenous dust clouds rotating across the stellar
disk.  To distinguish between these two possibilities it will be
necessary to re-observe in multiple filters (or with time resolved
spectroscopy) to measure the change in $T_{\rm eff}$, and in filters
sensitive to high dust opacity.

\begin{acknowledgements}
We would like to thank James Liebert and the 2MASS team for supplying
information on the 2MASS L dwarfs prior to publication. This work is
based on observations made with the 2.2m telescope at the
German--Spanish Astronomical Center at Calar Alto in Spain.
\end{acknowledgements}

\end{document}